\documentclass[10pt,letterpaper]{article}
\usepackage[top=0.85in,left=2.75in,footskip=0.75in]{geometry}

\usepackage{amsmath,amssymb}

\usepackage{changepage}

\usepackage[utf8x]{inputenc}

\usepackage{textcomp,marvosym}

\usepackage{cite}

\usepackage{nameref,hyperref}

\usepackage[right]{lineno}

\usepackage{microtype}
\DisableLigatures[f]{encoding = *, family = * }

\usepackage[table]{xcolor}

\usepackage{subcaption}
\usepackage{booktabs}
\usepackage{multirow}
\usepackage{multicol}

\usepackage{array}

\newcolumntype{+}{!{\vrule width 2pt}}

\newlength\savedwidth

\usepackage{setspace} 
\raggedright
\usepackage[aboveskip=1pt,labelfont=bf,labelsep=period,justification=raggedright,singlelinecheck=off]{caption}
\renewcommand{\figurename}{Fig}

\makeatletter
\renewcommand{\@biblabel}[1]{\quad#1.}
\makeatother

\usepackage{lastpage,fancyhdr,graphicx}
\usepackage{epstopdf}
\fancyheadoffset[L]{2.25in}
\fancyfootoffset[L]{2.25in}
\begin{document}
\vspace*{0.2in}

\begin{flushleft}
{\Large
\textbf\newline{A mathematical modelling framework for the regulation of intra-cellular OCT4 in human pluripotent stem cells} 
}
\newline
\\
L E Wadkin\textsuperscript{1*},
S Orozco-Fuentes\textsuperscript{2},
I Neganova\textsuperscript{3},
M Lako\textsuperscript{4},
N G Parker\textsuperscript{1},
A Shukurov\textsuperscript{1}
\\
\bigskip
\textbf{1} School of Mathematics, Statistics and Physics, Newcastle University, UK, NE1 7RU
\\
\textbf{2} Department of Mathematics, Physics and Electrical Engineering, Northumbria University, Newcastle upon Tyne, UK
\\
\textbf{3} Institute of Cytology, RAS St Petersburg, Russia
\\
\textbf{4} Bioscience Institute, Newcastle University, UK, NE1 3BZ
\bigskip

%
%





* l.e.wadkin@newcastle.ac.uk

\end{flushleft}
\section*{Abstract}
Human pluripotent stem cells (hPSCs) have promising clinical applications in regenerative medicine, drug-discovery and personalised medicine due to their potential to differentiate into all cell types, a property know as pluripotency. A deeper understanding of how pluripotency is regulated is required to assist in controlling pluripotency and differentiation trajectories experimentally. Mathematical modelling provides a non-invasive tool through which to explore, characterise and replicate the regulation of pluripotency and the consequences on cell fate. Here we use experimental data of the expression of the pluripotency transcription factor OCT4 in a growing hPSC colony to develop and evaluate mathematical models for temporal pluripotency regulation. We consider fractional Brownian motion and the stochastic logistic equation and explore the effects of both additive and multiplicative noise. We illustrate the use of time-dependent carrying capacities and the introduction of Allee effects to the stochastic logistic equation to describe cell differentiation. This mathematical framework for describing intra-cellular OCT4 regulation can be extended to other transcription factors and developed into sophisticated predictive models.



\section*{Introduction}
\label{sec:intro}

Human pluripotent stem cells, hPSCs, have the ability to self-renew through repeated divisions and to differentiate into a wide range of cell types, a property known as pluripotency. The pluripotency of hPSCs is their defining characteristic, central to their touted applications in drug discovery, regenerative and personalised medicine \cite{Ebert10,Zhu13,Avior16,Ilic17,Shroff17,Trounson16}. However, hPSCs exhibit complex behaviour and the \textsl{in-vitro} control of their differentiation trajectories is challenging. 

Pluripotency is controlled by an inter-regulatory network of pluripotency transciption factors, PTFs, including the genes OCT4, SOX2 and NANOG \cite{Li04,Young05,Chambers09}. The destabilisation of PTFs and their interaction with chemical signalling pathways result in differentiation away from the pluripotent state and into a specialised cell \cite{Li04,Kumar14,Wang12}. This decision of a cell to either remain pluripotent or to differentiate is known as its fate decision. It is unknown how much cell fate decisions are led by inherited factors, as opposed to environmental factors and intra-cellular signalling as even clonal (genetically identical) cells under apparently identical conditions make different fate decisions \cite{Symmons16}. In many \textsl{in-vitro} experiments the differentiation of hPSC populations is induced and facilitated by a differentiation agent, such as BMP4 \cite{Kee06,Xu02}.

A narrow range of PTF expression is necessary to maintain cell pluripotency, with both high and low expressions causing a shift from the pluripotent state \cite{Niwa00,Kopp08} and even small fluctuations can bias cell fate decisions \cite{Strebinger19}. Furthermore, the PTFs are inherited asymmetrically as a cell divides, biasing the fate of the daughter cells and contributing to colony heterogeneity \cite{Wolff18,Skamagki13,Tee14} with the decision to differentiate largely determined before any differentiation stimulus is introduced \cite{Wolff18}. Given the likely large number of factors involved in the fate decisions and our limited knowledge of their nature, the probabilistic framework to modelling PTF dynamics appears to be the most suitable. However, careful, experiment-based quantification of the stochastic, temporal dynamics of PTFs is necessary to examine the resulting effects on cell fate. 

Statistical analysis and mathematical modelling are deepening our understanding of hPSC behaviours and guiding the development of experimental protocols \cite{Wadkin20}. Recent mathematical models of cell pluripotency focus on describing the network of PTFs and the resulting cell fate decisions to guide the optimisation and control of pluripotency \textsl{in-vitro} \cite{Wadkin20,Herberg15,Pir16}. These models are informed by recent studies of fluctuations of PTFs throughout colonies \cite{Strebinger19,Wolff18,Torres14} and the spatial patterning of differentiation \cite{Rosowski15,Warmflash14}. Many models use coupled differential equations based on the Hill equations \cite{Hill1913} describing changes in concentrations of molecules to describe PTF fluctuations \cite{Glauche10,Chickarmane06,Akberdin18}. Others use network analysis frameworks \cite{Xu14} or explore the mechanical aspects of the cell behaviour when both the model and data are complex \cite{Auddya17}. These models often aim to describe the whole PTF regulatory network and it can be difficult to estimate the model parameters accurately from experimental data \cite{Akberdin18}. 

Here we focus on the methodology of building such mathematical models using experimental data for the transcription factor OCT4. Although the OCT4 dynamics will be affected by many external factors and the remainder of the PTF network, there are benefits to considering each PTF in isolation as the crucial first step; firstly, this simplifies the model development process, allowing each element to be explored in a systematic way and secondly, the results provide a basis for comparison to the other PTFs (e.g., NANOG and SOX2) from similar experiments. Similarly, although interesting spatial patterning effects are seen in OCT4 \cite{Wadkin20oct}, we will consider only the intra-cellular OCT4 behaviour through time. These simpler models can be used to describe the stochastic nature of PTF regulation on shorter time scales and explore the effects of each PTF on cell fate, before their development into coupled models of the entire pluripotency regulatory network.

Here we systematically explore various mathematical models for the temporal regulation of the PTF OCT4. We aim to identify the optimal set of mathematical tools required to reproduce the key quantitative features of experimental observations from Ref. \cite{Wolff18} and the additional quantitative analysis of this dataset from Ref. \cite{Wadkin20oct}. The framework discussed can be applied in future to other experimental datasets. Since PTF fluctuation is inherently stochastic \cite{Wolff18,Torres14,MacArthur16,Holmes17}, we focus on different forms of well-established stochastic models to describe the behaviour, namely: fractional Brownian motion and the stochastic logistic equation. The aim is to describe the PTFs as microstates before considering the macrostate of cellular pluripotency. Firstly, we introduce the experimental data and outline the key features of OCT4 to be described mathematically. Next, we explore fractional Brownian motion and the stochastic logistic equation for simulating temporal OCT4 before any cell differentiation occurs. We consider different types of random noise (additive and multiplicative \cite{Hasty00,liu2009effect}) and their effects. Finally, we examine the use of shifting carrying capacities and Allee effects to simulate a reduction in OCT4 towards the differentiated state. 

\section*{Experimental OCT4 fluctuations}
\label{sec:exp}

We use experimental data of OCT4 expression in a growing hESC colony from Ref. \cite{Wolff18} and our previous analysis of this data in Ref. \cite{Wadkin20oct} to guide model development. Although focused on one experiment, the mathematical framework outlined here is easily adaptable to other experimental results. We use the experimental analysis in Ref.~\cite{Wolff18} and Ref.~\cite{Wadkin20oct} to illustrate the applicability of such models to PTF regulation. Here we summarise the experiment and main features of the data to be described by a mathematical model. 

\subsection*{Experiment summary}
This experiment was carried out by Purvis Lab (University of North Carolina, School of Medicine), and is published in Ref. \cite{Wolff18}. The OCT4 levels (mean OCT4-mCherry fluorescence intensity) in a human embryonic stem cell colony were determined and cells were live-imaged for 68 hours. The colony begins from 30 cells and grows over 68\,hours (817 time frames) to 463 cells, with 1274 cell cycles elapsing within this time. After 40 hours, the hESCs were treated with (100\,ng/ml) bone-morphogenetic protein 4 (BMP4) to induce their differentiation towards distinct cell fates. The cell IDs, ancestries and positions, ($x(t)$, $y(t)$), were extracted along with their OCT4 immuno-fluorescence intensity values (reported in arbitrary fluorescence units, a.f.u.). The measurements of the OCT4 signal at 5 minute intervals, results in a set of evenly sampled discrete observations for each cell, OCT4$(t_0)$, OCT4$(t_1), ..., $ OCT4$(t_n)$, where $t_0$ and $t_n$ denote the times of cell birth and division, respectively. The values of $t_n$ range from 0.25--30 hours across the population.

To classify the cells as either self-renewing (pluripotent) or differentiated, the expression levels of CDX2 were quantified at 68 hours. The final cells were fate classified using CDX2 and OCT4 as either belonging exclusively to a pluripotent or differentiated state. A remaining group of cells were classified in an unknown category. Using these fates, the cell population was traced back in time, spanning multiple cell divisions, with each earlier cell labelled according to this pro-fate. In this paper we consider only the pluripotent and differentiated fate groups. Note that for times pre-BMP4 (before 40 hours), the fate classification is a pro-fate based on the fate of the cells descendants. 

\subsection*{Temporal OCT4 features}
\label{sec:features}
\begin{figure}[!h]
	\includegraphics[width=\textwidth]{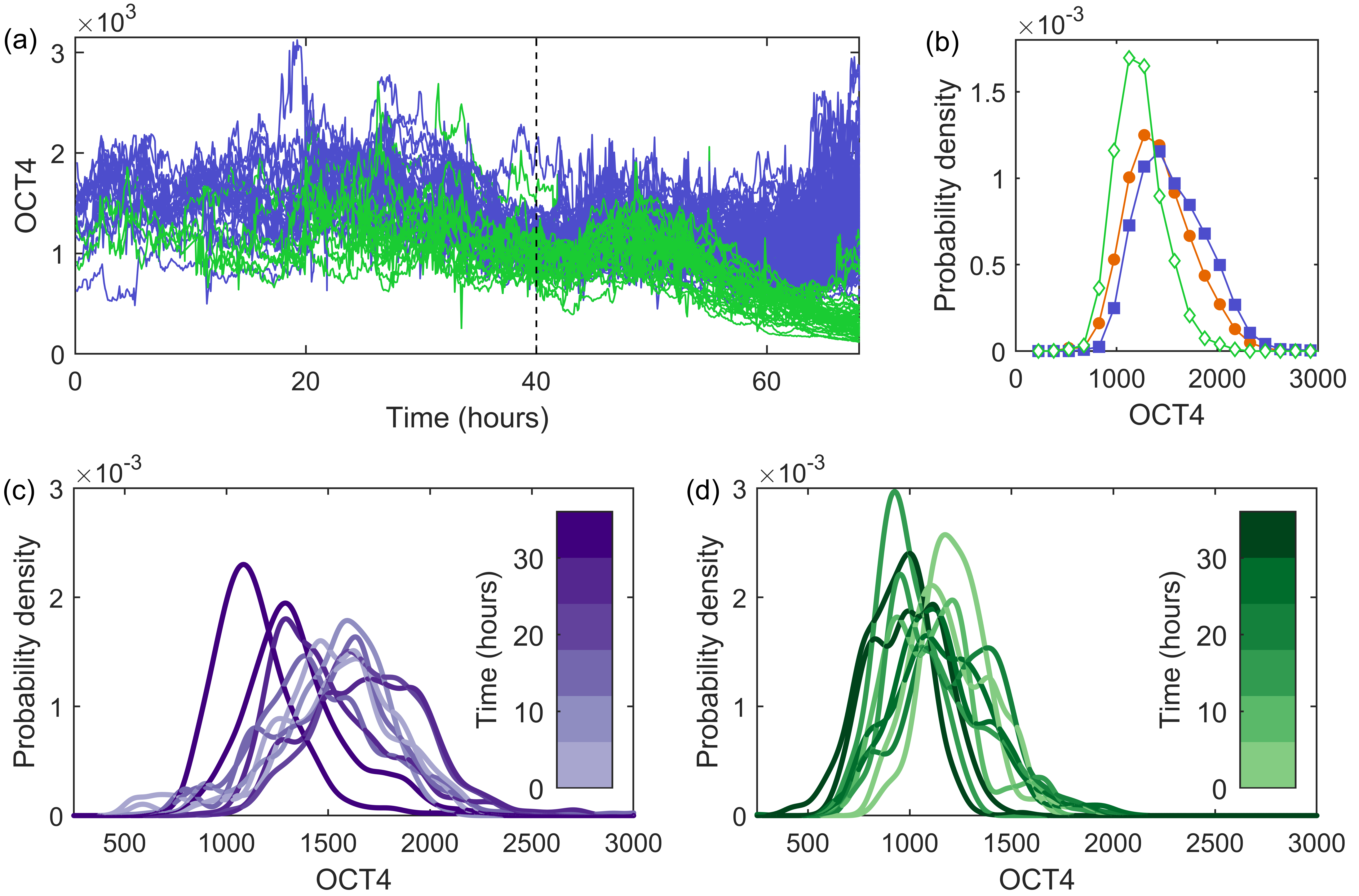}
	\caption{{\bf Experimental OCT4 properties.}
		(a) The temporal OCT4 expression for all (pro-)pluripotent (purple) and (pro-)differentiated (green) cells up to 68 hours. At 40 hours (dashed line) the differentiation agent BMP4 was added. Pre-differentiation: (b) The distribution of all OCT4 expressions for all (orange circles), pro-pluripotent (purple squares) and pro-differentiated (green diamonds) cells. The distribution of OCT4 expression for time intervals between zero and 40\, hours for (c) pro-pluripotent and (d) pro-differentiated cells from the experiment. The colour bar shows the time of the bin centre.}
	\label{fig:octovertime}
\end{figure}

The OCT4 expression of (pro-)pluripotent and (pro-)differentiated cells for the whole experimental time (68 hours) is shown in Fig~\ref{fig:octovertime}(a). At 40 hours the differentiation agent BMP4 is added, after which there is a decline in OCT4 expression in the (pro-)differentiated cells. The (pro-)pluripotent cells retain their OCT4 expression levels. The distribution of all OCT4 expressions pre-differentiation is shown in Fig~\ref{fig:octovertime}(b), with temporal distributions in Fig~\ref{fig:octovertime}(c) and (d) for pluripotent and differentiated pro-fate cells respectively. A detailed analysis of the experimental data is provided in Ref. \cite{Wadkin20oct}. We identify several key features to capture in model development, summarised below. For simplicity, and due to the distinct behavioural differences identified pre- and post-differentiation, we first consider modelling the temporal behaviour pre-BMP4 before moving on to the effect of cell differentiation.

\begin{itemize}
	\item Pre-differentiation
	\begin{enumerate}
		\item{The time series exhibit stochastic noise with the Hurst exponent 0.38 in both (pro-)pluripotent and (pro-)differentiated cells, shown in Fig~\ref{fig:octovertime}(a) and calculated in Ref. \cite{Wadkin20oct}. A Hurst exponent $<0.5$ indicates anti-persistence in the time series, with increases in OCT4 more likely to be followed by decreases, and vice versa.}
		\item{Pro-differentiated cells show reduced OCT4 expression throughout, shown in Fig~\ref{fig:octovertime}(a) and (b).}
		\item{The distribution of all OCT4 expressions from (pro-)pluripotent cells is positively skewed, resulting from a reduction in expression at later times, shown in Fig~\ref{fig:octovertime}(b) and (c).}
		\item{The distribution of all OCT4 expressions from (pro-)pluripotent cells show a temporal shift in the mode, with a reduction in expression with time, shown in Fig~\ref{fig:octovertime}(c).}
	\end{enumerate} 
	
	\item Post-differentiation 
	\begin{enumerate}
		\item At the end of the experiment differentiated cells are classified according to their OCT4 and CDX2 expressions. These differentiated cells shown a pronounced reduction in OCT4 upon BMP4 addition (40 hours).
		\item There is a clear and natural separation between the two classified groups post-BMP4 based on their OCT4 levels, with differentiated cells showing reduced OCT4 and pluripotent cells retaining OCT4 expression. 
	\end{enumerate}
\end{itemize}

\section*{Results}
\subsection*{Modelling OCT4 pre-differentiation}
\label{sec:modelling}
In the following sections we systematically explore the use of different stochastic models as a framework for temporal OCT4 regulation, aiming to capture the experimental behaviour described above and shown in Fig~\ref{fig:octovertime}. All the models discussed have the same basis, with the initial conditions and cellular division incorporated using the algorithmic base model detailed below.

\subsubsection*{Base Model}
\begin{figure}[!h]
	\includegraphics[width=\textwidth]{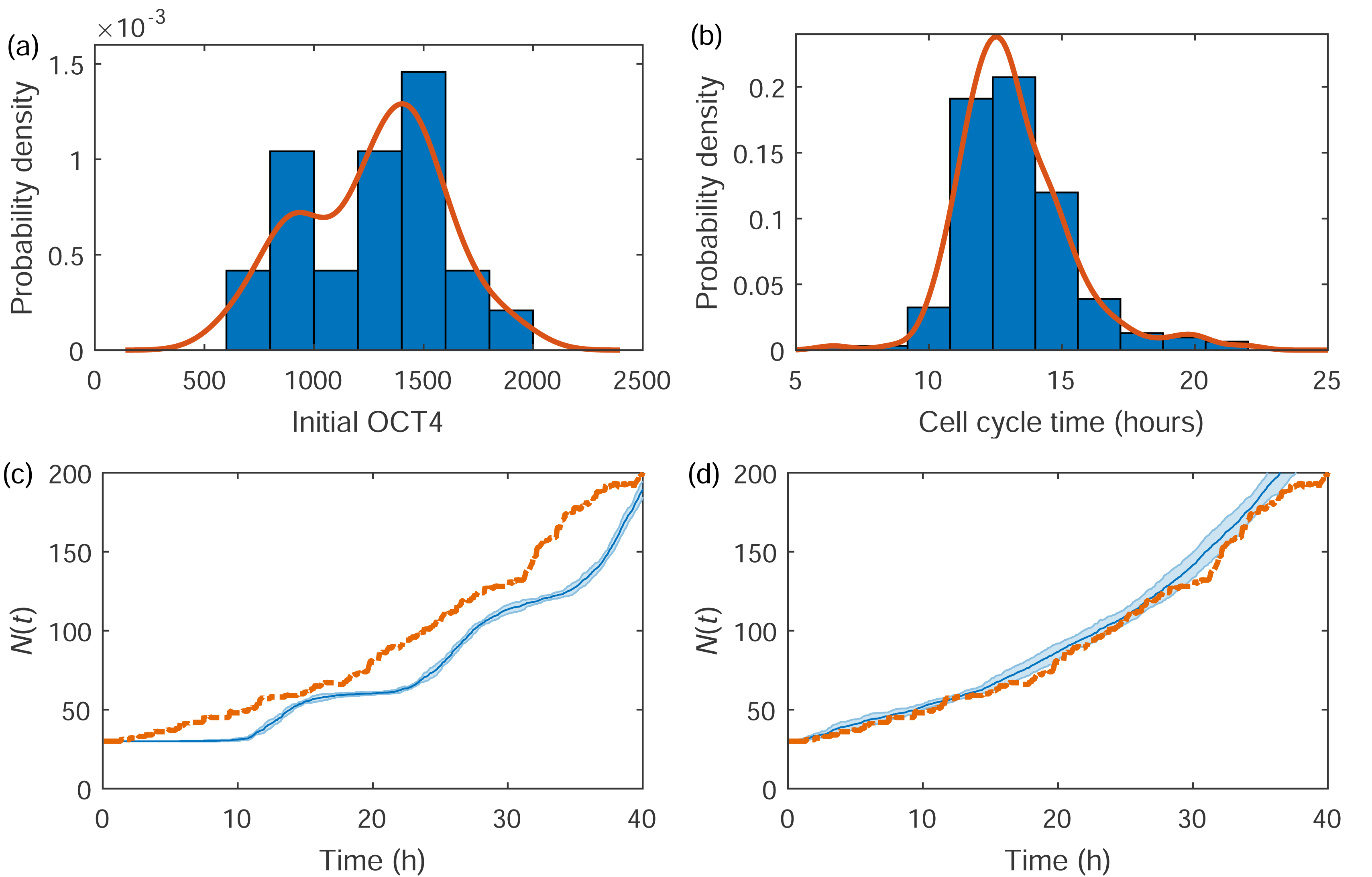}
	\caption{\label{fig:basemodel}{\bf The initial conditions and resulting population dynamics for the common base model.} (a) The distribution of experimental initial OCT4 values histogram, OCT4$(t=0)$, with kernel density fitting shown in orange. (b) The experimental distribution of cell cycle duration times histogram for all cells pre-BMP4 addition with kernel density fitting shown in orange. The number of cells over time, $N(t)$, when cellular division is (c) synchronised and (d) not synchronised in step 3 of the common base model. Blue solid lines show the simulated population sizes with standard deviation error range shown in light-blue (calculated over five realisations) and orange dashed lines show the experimental population.}
\end{figure}

\begin{enumerate}
	\item We begin with a chosen initial number of cells, $N=N_0$, to match the experimental conditions.
	\item Each of the $N$ cells are allocated an initial OCT4 value. This is extracted probabilistically from the kernel density fitting to the experimental distribution of initial  OCT4, OCT4$(t=0)$, shown in Fig~\ref{fig:basemodel}(a).
	\item Each of the $N$ cells are allocated a cell cycle duration. This is extracted probabilistically from the kernel density fitting to the experimental distribution of cell cycle times for all pre-BMP4 cells, shown in Fig~\ref{fig:basemodel}(b). Each cell's starting position in its cell cycle is chosen uniformly.
	\item For each of the $N$ cells the OCT4 values for the duration of their cell cycle are simulated using one of the stochastic models. 
	\item Each of the $N$ cells divide into two cells at the end of their cell cycle. For each of the two daughter cells, their initial OCT4 value is set to the pre-division OCT4 value of the mother cell. 
	\item Steps 4 and 5 are repeated for the number of required division events.
\end{enumerate}

When the cell cycle times are generated in step 3 it is necessary to specify how much of the cell cycle has already elapsed. If all cells begin at the start of their cell cycle at the start of the simulation then divisions will be synchronised, visible as `steps' in the number of cells over time, $N(t)$, as shown in Fig~\ref{fig:basemodel}(c). Avoiding this synchronisation by starting cells at different points in their cell cycle gives a more accurate representation of colony size, as shown in Fig~\ref{fig:basemodel}(d). 

Although here we have used the analysis of the experimental data to inform the initial conditions and the cell cycle simulation, this is flexible and can easily be adapted to other experimental results. The OCT4 regulation itself is captured in step 4 and is open to many mathematical modelling techniques. In the next section we use the experimental results from Ref.~\cite{Wolff18} and \cite{Wadkin20oct} to systematically build a stochastic model using fractional Brownian motion and the stochastic logistic equation.

\subsubsection*{Anti-persistent OCT4 fluctuations}
One possibility for a simple model of OCT4 fluctuation is to assume that the expression fluctuates symmetrically with no preferred trends or correlations. Mathematically this would be descried by a Wiener process, analogous to the physical phenomenon of Brownian motion in one dimension and the starting point for many random walk models. However, the analysis of experimental OCT4 expression described above and in Ref. \cite{Wadkin20oct} has shown that the OCT4 evolution is anti-persistent, with an average Hurst exponent of $H=0.38$. This signifies that increases in OCT4 are more likely to be followed by decreases, and vice versa. The Hurst exponent $H\neq 0.5$ indicates that the fluctuations in OCT4 cannot be captured by simple Brownian motion. 

Instead we consider the generalisation, fractional Brownian motion (fBm). Unlike Brownian motion, fBm allows for non-independent increments and hence persistence or anti-persistence. An fBM random function of time $t$, $B_H(t)$, with an initial value $B_H(0)$ and time increments $B_H(t-s)$ is defined by
\begin{multline}
B_H(t)=B_H(0)+\frac{1}{\Gamma(H+0.5)}\int_{-\infty}^{0}\left[(t-s)^{H-0.5}-(-s)^{H-0.5}\right]dB(s)\\+\frac{1}{\Gamma(H+0.5)}\int_{0}^{t}(t-s)^{H-0.5}dB(s),
\label{eq:fbm}
\end{multline}
where $H$ is the Hurst exponent and $\Gamma$ is the gamma function \cite{Mandelbrot68}. There are several ways to simulate fBm, either exact or approximate \cite{Dieker03,Dieker04,Yin96}. Here we use the Matlab function \textit{ffgn} \cite{ffgn} which uses the circulant embedding technique for $H<0.5$ \cite{Dietrich97} and Lowen's method \cite{Lowen99} for $H>0.5$ (both exact methods) to simulate the fractional Brownian noise. There is also an inbuilt Matlab function \textit{wfbm} (available in the Wavelet toolbox) which uses a wavelet based approximate simulation method \cite{Abry96}.

We can use fBm to simulate OCT4 over time (step 4 of the base model) with a scaling parameter $\sigma$ which controls the level of noise, i.e., $\sigma B_H$. Example realisations of the fractional noise, corresponding fBm functions, and simulated OCT4  for varying $H$ are shown in \nameref{fbm} to illustrate the effect of the Hurst exponent. The parameter $\sigma$ is estimated from the experimental data (for all pre-BMP4 cells) as the standard deviation of $\Delta\textrm{OCT4}=\textrm{OCT4}(t)-\textrm{OCT4}(t-1)$, leading to $\sigma\approx90$. Each time series for OCT4 can then be generated as OCT4$(t=0)+\sigma B_H$.

For simplicity, we first consider both cell fates together with $N=16$ cells, made up of 14 pro-pluripotent and two pro-differentiated cells to correspond to the experimental data \cite{Wolff18}. For cells in the experimental colony $H=0.38$ \cite{Wadkin20oct}. A comparable simulation using fBm with 16 initial cells, $H=0.38$, and $\sigma=90$ is shown in \nameref{fbmsim}. Note that although we simulate from a limited number of starting cells, the number of OCT4 values generated over 40 hours due to the 5 minute increments and cellular division is approximately 30000. It is clear from \nameref{fbmsim} that this level of anti-persistent regulation from the Hurst exponent is not sufficient to keep the OCT4 expression within the range seen in the experiment. 

One possible method of limiting the range of OCT4 is to impose boundary conditions, such as absorbing or reflecting. For absorbing boundary conditions once the OCT4 level reaches the boundary the cell is theoretically removed in some way from the experiment and its OCT4 time series does not continue. There is no indication or biological evidence of particularly high or low OCT4 expressions resulting in cell death experimentally \cite{Wolff18,Wadkin20oct}. However, high or low OCT4 expressions do accompany cell differentiation \cite{Strebinger19}, so the removal of cells via the boundary condition could correspond to the differentiation of cells if we consider pluripotent cells only. The OCT4 simulation for fBm with absorbing boundary conditions is shown in \nameref{fbmsim}.

Reflecting boundary conditions imply that when the OCT4 expression reaches the boundary, it is reflected back in the opposite direction. Biologically this corresponds to an additional regulatory effect, either internal to the cell or external in experimental conditions; if the OCT4 level in a cell becomes too low, there is systematic regulation to increase it (and vice versa). The simulation using fBm with reflecting boundary conditions is shown in \nameref{fbmsim}. Reflecting boundary conditions produce a result more similar to the experiment than absorbing boundary conditions since cells are not artificially removed, but it still creates a sharper distribution boundary than seen experimentally. Additionally, although the boundary conditions somewhat artificially force the OCT4 into the desired range, the spread of the overall expressions is not well captured.

This illustrates that the anti-persistence from the Hurst exponent alone is not sufficient to capture the OCT4 regulation seen in the experiment, even with boundary conditions. The imposition of any boundary conditions also requires further investigation to elucidate their nature, positioning and the biological implications. However, we can still incorporate fBm noise into other models to generate the anti-persistence. In the next section we consider describing temporal OCT4 with the stochastic logistic equation and explore the regulatory effects of a limiting carrying capacity.

\subsubsection*{The stochastic logistic equation}
\label{sec:stochlog}

In this section we explore the application of the stochastic logistic equation (SLE) to simulating temporal OCT4 regulation. The logistic equation is a widely used model of population dynamics characterized by the growth rate of the population and its optimal size called the carrying capacity. We adapt the logistic equation to the experimental data available, using the model for OCT4 variation, rather than the traditional population size. Since fBm alone does not fully capture the regulatory behaviour of OCT4, some additional effects are clearly important. We consider the SLE with additive noise, multiplicative noise, and the effect of a time-dependent carrying capacity. For simplicity, we again consider the two cell fates together initially.

There are several ways stochasticity can be introduced into the logistic equation, e.g., additive noise, multiplicative noise, a noisy growth rate parameter $r$ or carrying capacity $K$. Both additive and multiplicative noise can be used to regulate gene expression \cite{Hasty00}. The most straightforward of these is additive noise which can be introduced by adding a noise term to the net rate of change in the PTF. The SLE with additive random scatter to describe OCT4, $O$, over time, $t$, is then 
\begin{equation}
\frac{dO}{dt}=rO\left(1-\frac{O}{K}\right)+\sigma_{\rm{A}} \xi,
\label{eq:stochlog2}
\end{equation} 
where $\xi$ is the stochastic noise (e.g., Wiener/Brownian noise, or fBM noise) and $\sigma_{\rm{A}}$ is a scaling parameter controlling the magnitude of the scatter.

\begin{figure}[!h]
	\includegraphics[width=\textwidth]{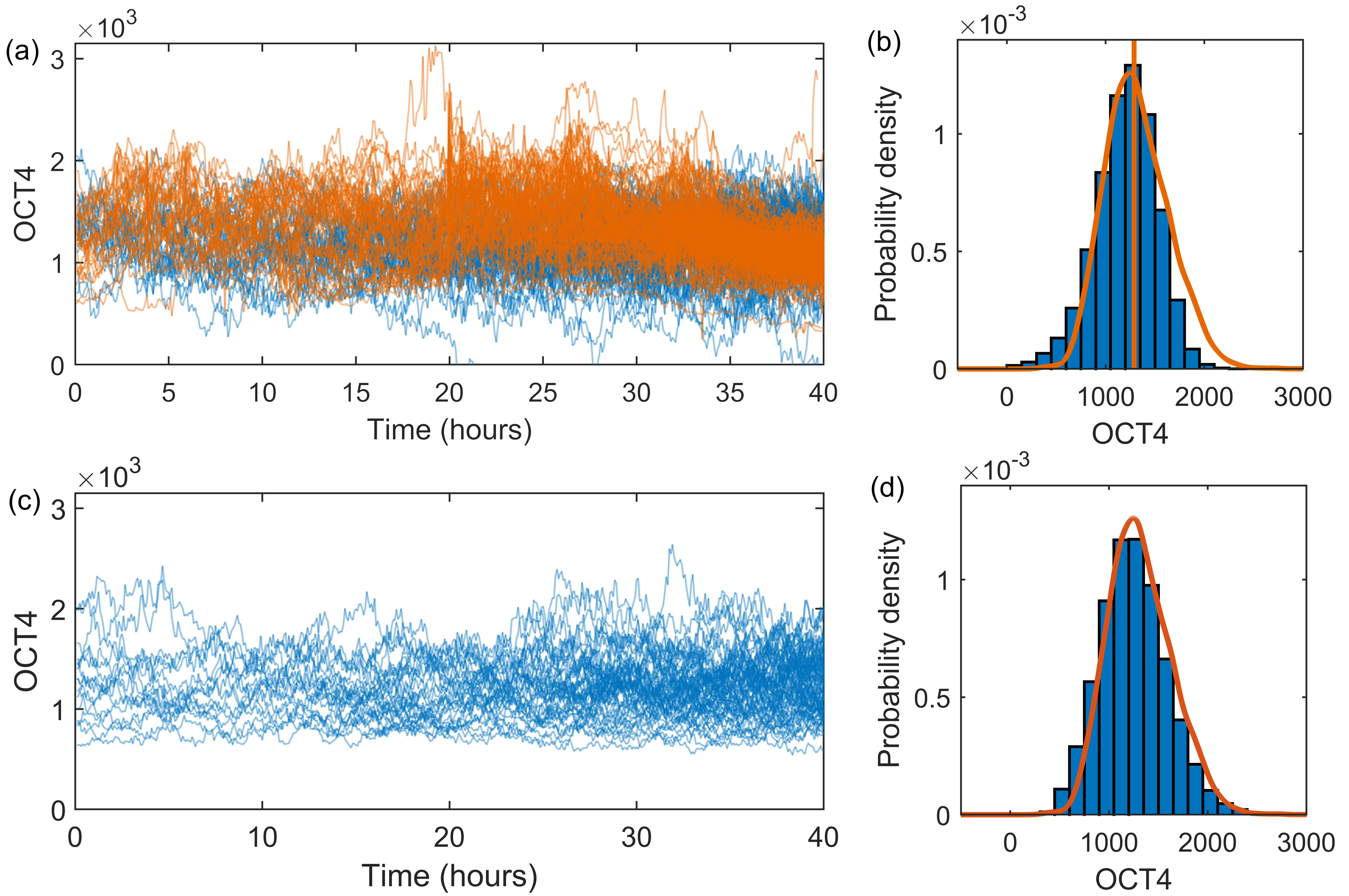}
	\caption{\label{fig:sle_addnoise}{\bf Comparison of experimental and simulated OCT4 using the SLE with either additive or multiplicative noise.} (a) Simulated OCT4 expression (blue) using the SLE with additive noise, Eq~(\ref{eq:stochlog2}), with 16 initial cells, $r=0.02$, $K=1290$, $\sigma_{\rm{A}}=90$ and fBM noise with $H=0.38$, with an absorbing boundary condition at zero. The experimental OCT4 is shown in orange. (b) The corresponding histogram of simulated OCT4 expression using Eq~(\ref{eq:stochlog2}) with the experimental distribution and estimated carrying capacity ($K=1290$) in orange. (c)~Simulated OCT4 expression using the SLE with multiplicative noise, Eq~(\ref{eq:stochlog2mult}), with 16 initial cells, $r=0.005$, $K=1290$, $\sigma_{\rm{M}}=0.0045$ and fBM noise with $H=0.38$. (d) The corresponding histogram of simulated OCT4 expression with the experimental distribution in orange.}
\end{figure}

Parameters that appear in Eq~(\ref{eq:stochlog2}) are estimated from the experimental data (pre-BMP4). In keeping with the anti-persistence, the noise $\xi$ corresponds to fBm noise with the Hurst exponent $H=0.38$ and the scaling parameter is again the standard deviation of $\Delta$OCT4, $\sigma_{\rm{A}}=90$. We can also estimate the carrying capacity as the median of all the experimental OCT4 values, $K=1290$. The OCT4 dynamics simulated using Eq~(\ref{eq:stochlog2}) with $r=0.02$ is illustrated in Fig~\ref{fig:sle_addnoise}(a) and (b). Although the regulatory effect of the carrying capacity works well to capture the upper bound of OCT4 expression, an additional boundary condition at small values of OCT4 is still required (if the stochasticity gives rise to $O<0$ then $dO/dt<0$ resulting in $O\to-\infty$). A distinguishing feature not captured by the model is the positive skew in the distribution of all occurring OCT4 values, shown in Fig~\ref{fig:octovertime}(b) and overlaid in Fig~\ref{fig:sle_addnoise}(b). The model promotes tighter regulation above the carrying capacity than below it, resulting in fewer OCT4 expressions above the carrying capacity than seen experimentally. This suggests that the stochasticity has some dependence on the current state of the system.

Whereas the additive noise in Eq~(\ref{eq:stochlog2}) has no dependence on the state of the system and corresponds to making $dO/dt$ symmetrically noisy, multiplicative noise changes depending on the current conditions. In the case of our temporal OCT4 simulation, multiplicative noise can be used to generate a scatter in the simulated data which has a greater magnitude when the system is close to the carrying capacity (thus resulting in more stochastically high OCT4 expressions) and a reduced magnitude when far away from the carrying capacity. Hints of this behaviour can be seen in Fig~\ref{fig:octovertime}(a), with larger fluctuations apparent in the cells exhibiting above average OCT4 expression. For simulating the SLE with multiplicative noise we first consider the rearrangement of the logistic equation,
\begin{equation*}
\frac{d\ln(O)}{dt}=r\left(1-\frac{O}{K}\right).
\end{equation*}
Applying the substitution $X=\ln(O)$ and adding stochasticity $\xi$ with noise scaling parameter $\sigma_{\rm{M}}$ gives
\begin{equation}
\frac{dX}{dt}=r\left(1-\frac{e^X}{K}\right)+\sigma_{\rm{M}}\xi,
\label{eq:stochlog2mult}
\end{equation}
which can then be used to simulate $X=\ln(O)$, with the dynamics of OCT4 recovered from $O=e^X$. Example realisations of Eq~(\ref{eq:stochlog2mult}) for both $X$ and $O$ are shown in \nameref{loglogisticeg} to illustrate the effect of multiplicative noise in a typical logistic growth scenario for varying $\sigma_{\rm{M}}$. The result is amplified noise for stochasticity occurring above the carrying capacity.

The temporal OCT4 dynamics simulated using the SLE with multiplicative noise, Eq~(\ref{eq:stochlog2mult}), with fBM noise with $H=0.38$, $r=0.005$, $K=1290$ and $\sigma_{\rm{M}}=0.0045$ for 16 initial cells are shown in Fig~\ref{fig:sle_addnoise}(c). The multiplicative noise results in cells with expressions above the carrying capacity exhibiting increased stochasticity, with lower expression cells showing tighter regulation. The simulated distribution has a slight positive skew and is qualitatively similar to the experimental distribution, as shown in Fig~\ref{fig:sle_addnoise}(d). 

This model provides a good basis for capturing the experimental results across the whole time period and is an improvement on the SLE with additive noise. However, it does not take into account the different cell fates, and the evolving temporal positive skew in the pluripotent cell group, shown in Fig~\ref{fig:octovertime}(c). In the following sections we consider the two cell fates separately and discuss two methods of including the temporal skew in the pluripotent cell group: the SLE with a transition between dominant additive and dominant multiplicative noise, and the SLE with a time-dependent carrying capacity. 

\subsubsection*{SLE with noise transition}

Firstly, to capture the changing temporal skew for pluripotent cells, we could include both additive and multiplicative noise because different noise types reflect different aspects in the cell behaviour \cite{liu2009effect} and both appear to be involved in the experimentally observed evolution of OCT4. If additive noise is dominant at early times, and multiplicative noise at later times, the resulting OCT4 distribution will be symmetric at early times and skewed at later times. We can consider the following rearrangement of the stochastic logistic equation with additive noise
\begin{equation*}
\frac{d\ln(O)}{dt}=r\left(1-\frac{O}{K}\right)+\frac{\sigma_{\rm{A}}}{O}\xi_1,
\end{equation*}
make the substitution $X=\ln(O)$ and introduce the multiplicative noise term $\sigma_{\rm{M}}\xi_2$,
\begin{equation}
\frac{dX}{dt}=r\left(1-\frac{e^X}{K}\right)+\frac{\sigma_{\rm{A}}}{e^X}\xi_1+\sigma_{\rm{M}}\xi_2.
\label{eq:stochlog2bothnoise}
\end{equation}
As before, we can simulate the dynamics for $X$ and recover the dynamics for $O=e^X$. 

\begin{figure}[!h]
	\includegraphics[width=\textwidth]{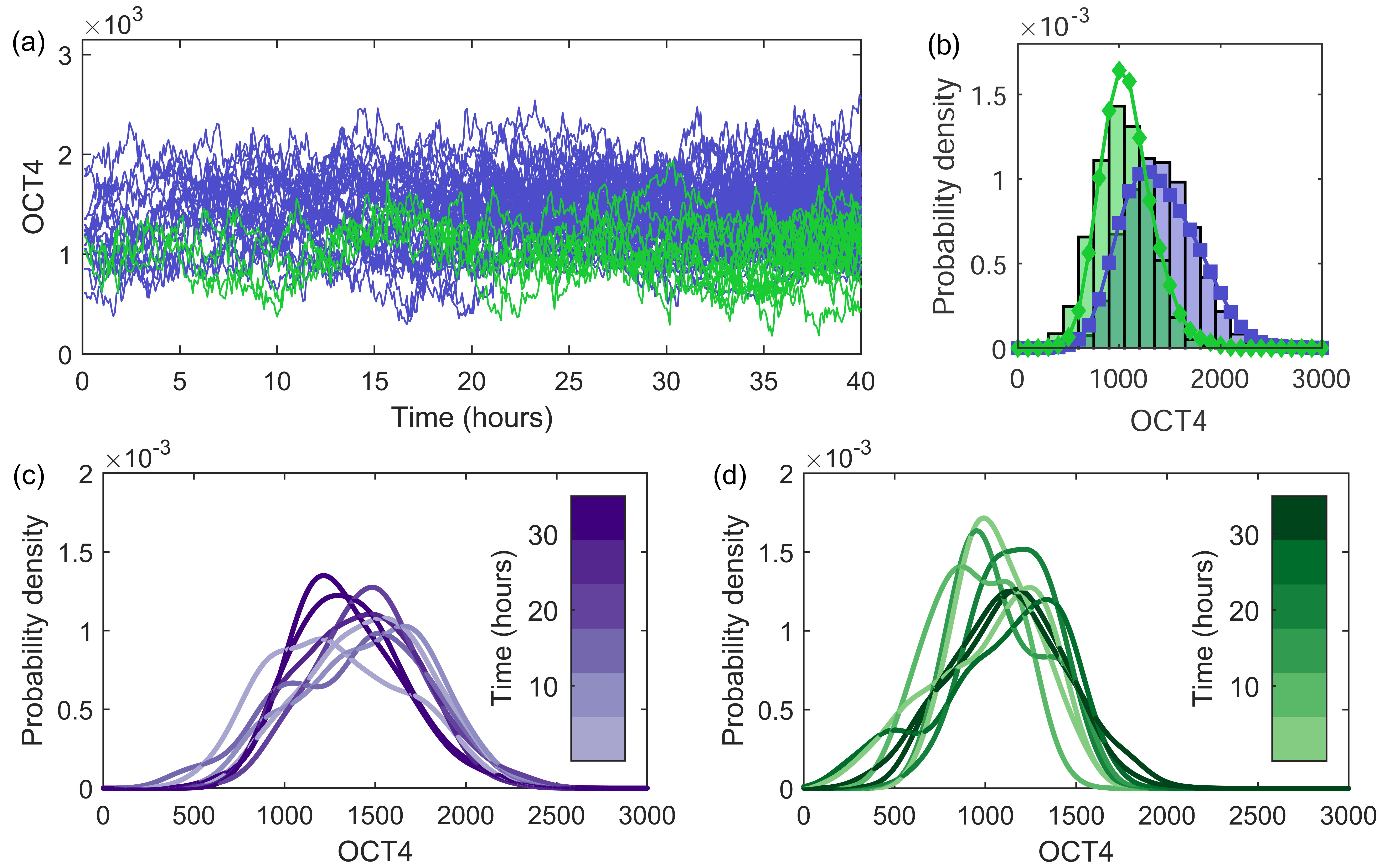}
	\caption{\label{fig:noiseswitch} {\bf The dynamics of OCT4 simulated using the SLE with a switch between additive and multiplicative noise.} (a) The OCT4 dynamics between zero and 40 hours for 14 pro-pluripotent (purple) and two pro-differentiated (green) initial cells following the SLE with both additive and multiplicative noise, Eq~(\ref{eq:stochlog2bothnoise}), with the parameters specified in Table~\ref{tab:sim_noiseswitch_parameters}. For pro-pluripotent cells the noise changes from additive to multiplicative at 20 hours. (b) The distribution of all simulated OCT4 values for pro-pluripotent (purple) and pro-differentiated (green) cells with the corresponding experimental distributions overlaid. The temporal distributions for (c) pro-pluripotent and (d) pro-differentiated cells split by time intervals.}
\end{figure}

For simplicity, we can consider the change between additive and multiplicative noise as a switch for pluripotent cells: for $0<t<20\,$h, $\sigma_{\rm{A}}=90$ and $\sigma_{\rm{M}}=0$, and for $t>20\,$h, $\sigma_{\rm{A}}=0$ and $\sigma_{\rm{M}}=0.05$. The additional parameters are specified in Table~\ref{tab:sim_noiseswitch_parameters}. Since differentiated cells show reduced OCT4 expression throughout, they are given a lower carrying capacity. The results for the OCT4 dynamics within this regime are shown in Fig~\ref{fig:noiseswitch}. The reduced carrying capacity for differentiated cells results in their lower expression throughout, shown in Fig~\ref{fig:noiseswitch}(a). The overall OCT4 expression distributions in Fig~\ref{fig:noiseswitch}(b) are well described. The temporal distributions in Fig~\ref{fig:noiseswitch}(c) illustrate the effect of the noise switch in the pluripotent cells, with the appearance of a positive skew at later times, while the expression of differentiated cells in Fig~\ref{fig:noiseswitch}(d) remains symmetrical at later times. 
\begin{table}[!h]
	\centering
	\caption{Simulation parameters for the OCT4 expression for pluripotent and differentiated cells using the SLE with both multiplicative and additive noise, Eq~(\ref{eq:stochlog2bothnoise}). At 20 hours the noise switches from additive to multiplicative noise in the pluripotent cells.}
	\begin{tabular}{@{}lccc@{}}
		\toprule
		& Parameter & $t<20\,$h & $t\geq20\,$h\\
		\midrule
		\multirow{5}{*}{Pluripotent} & $N_0$ & \multicolumn{2}{c}{14} \\
		& $r$ &  \multicolumn{2}{c}{0.01}\\
		& $K$ & \multicolumn{2}{c}{1290} \\
		& $\sigma_{\rm{A}}$ &  90 & 0 \\
		& $\sigma_{\rm{M}}$ & 0 & 0.05 \\
		& $H$ &  \multicolumn{2}{c}{0.38} \\
		\midrule
		\multirow{5}{*}{Differentiated} & $N_0$ &\multicolumn{2}{c}{2} \\
		& $r$ & \multicolumn{2}{c}{0.01}\\
		& $K$ & \multicolumn{2}{c}{1000} \\
		& $\sigma_{\rm{A}}$ & \multicolumn{2}{c}{90}\\
		& $\sigma_{\rm{M}}$ & \multicolumn{2}{c}{0} \\
		& $H$ & \multicolumn{2}{c}{0.38} \\
		\bottomrule
	\end{tabular}
	\label{tab:sim_noiseswitch_parameters}
\end{table}

\newpage
Although this model captures the overall distribution and provides the desired temporal change in skew (which could be further smoothed with a more sophisticated time-dependent noise function), it does not result in a shift in the mode expression as drastic as the one apparent in Fig~\ref{fig:octovertime}(c). For this we consider implementing a time-dependent carrying capacity in the next section. 

\subsubsection*{SLE with time-dependent carrying capacity}
To reproduce the significant shift in the mode for the pluripotent cells we can employ a time-dependent carrying capacity. We use the stochastic logistic equation for all cells, with both multiplicative and additive noise, as in Eq~(\ref{eq:stochlog2bothnoise}), and a carrying capacity which varies with time, 
\begin{equation}
\frac{dX}{dt}=r\left(1-\frac{e^X}{K(t)}\right)+\frac{\sigma_{\rm{A}}}{e^X}\xi_1+\sigma_{\rm{M}}\xi_2.
\label{eq:timedepk}
\end{equation}

\begin{figure}[!h]
			\includegraphics[width=\textwidth]{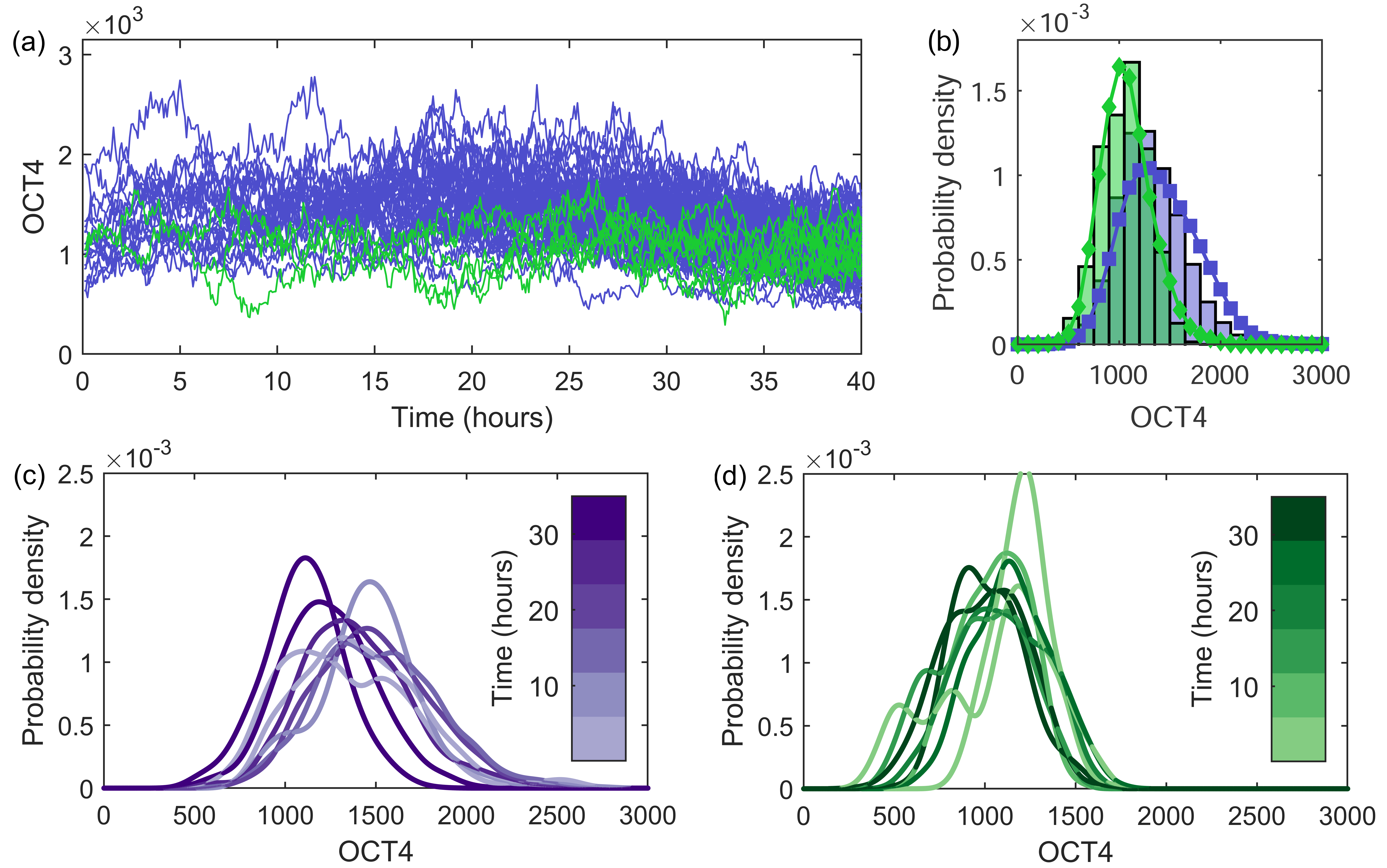}
	\caption{\label{fig:shiftk} {\bf The dynamics of OCT4 simulated using the SLE with a time-dependent carrying capacity.} (a) The OCT4 dynamics between zero and 40 hours for 14 pro-pluripotent (purple) and two pro-differentiated (green) initial cells following the SLE with both additive and multiplicative noise and a time-dependent carrying capacity, Eq~(\ref{eq:timedepk}), with the parameters specified in Table~\ref{tab:sim_timedepk_parameters}. For pro-pluripotent cells the carrying capacity reduces at 20 hours, whilst the carrying capacity for pro-differentiated cells is constant. (b) The distribution of all simulated OCT4 values for pro-pluripotent (purple) and pro-differentiated (green) cells with the corresponding experimental distributions overlaid. The temporal distributions for (c) pro-pluripotent and (d) pro-differentiated cells split by time intervals.}
\end{figure}

We can estimate the carrying capacity as the median OCT4 between zero and 25 hours resulting in $K_{\rm{p}}\approx$ 1500 and $K_{\rm{d}}\approx 1100$ for pluripotent and differentiated cells, respectively. For simplicity, post-25 hours, we will estimate both carrying capacities as $K\equiv K_{\rm{p}}=K_{\rm{d}}\approx 1000$. This reduction in the carrying capacity will initiate the corresponding reduction in the mode of the distribution over time we see experimentally. The OCT4 dynamics using the time-dependent carrying capacities in Eq~(\ref{eq:timedepk}) for 14 pro-pluripotent and two pro-differentiated cells, with the model parameters summarised in Table~\ref{tab:sim_timedepk_parameters} are shown in Fig~\ref{fig:shiftk}.

\begin{table}[!]
	\centering
		\caption{Simulation parameters for generating OCT4 expression for pro-pluripotent and pro-differentiated cells using the SLE with additive and multiplicative noise, and a time-dependent carrying capacity, Eq~(\ref{eq:timedepk}).}
	\begin{tabular}{@{}lccc@{}}
		\toprule
		& Parameter & $t<25\,$h & $t\geq25\,$h\\
		\midrule
		\multirow{5}{*}{Pluripotent} & $N_0$ & \multicolumn{2}{c}{14} \\
		& $r$ &  \multicolumn{2}{c}{0.015}\\
		& $K$ & 1500 & 1000 \\
		& $\sigma_{\rm{A}}$ &  \multicolumn{2}{c}{30} \\
		& $\sigma_{\rm{M}}$ & \multicolumn{2}{c}{0.035}\\
		& $H$ &  \multicolumn{2}{c}{0.38} \\
		\midrule
		\multirow{5}{*}{Differentiated} & $N_0$ &\multicolumn{2}{c}{2} \\
		& $r$ & \multicolumn{2}{c}{0.015}\\
		& $K$ & 1100 & 1000 \\
		& $\sigma_{\rm{A}}$ & \multicolumn{2}{c}{20}\\
		& $\sigma_{\rm{M}}$ & \multicolumn{2}{c}{0.03} \\
		& $H$ & \multicolumn{2}{c}{0.38} \\
		\bottomrule
	\end{tabular}
	\label{tab:sim_timedepk_parameters}
\end{table}

The lower carrying capacity results in consistently lower OCT4 expression for the differentiated cells, as shown in Fig~\ref{fig:shiftk}(a) and (b). The overall distribution of OCT4 expressions is well described, shown in Fig~\ref{fig:shiftk}(b). The model captures the shift to lower OCT4 values in pluripotent cells, shown in the temporal distribution in Fig~\ref{fig:shiftk}(c). The parameter choice could be further refined to additionally capture the change in the temporal skew using time-dependent multiplicative noise.

Here we have outlined some possible techniques for simulating temporal OCT4 using the SLE with different modes of fBm stochasticity and a time-dependent carrying capacity. Note that we aim to illustrate the application of such a model and describe a framework which could be used to capture some of the global properties of experimental data sets. Further work is now required to elucidate the appropriate parameter choices with further experiments and explore their biological implications.  

\subsection*{Simulating cell differentiation}
\label{sec:diff}
\begin{figure}[!h]
	\includegraphics[width=\textwidth]{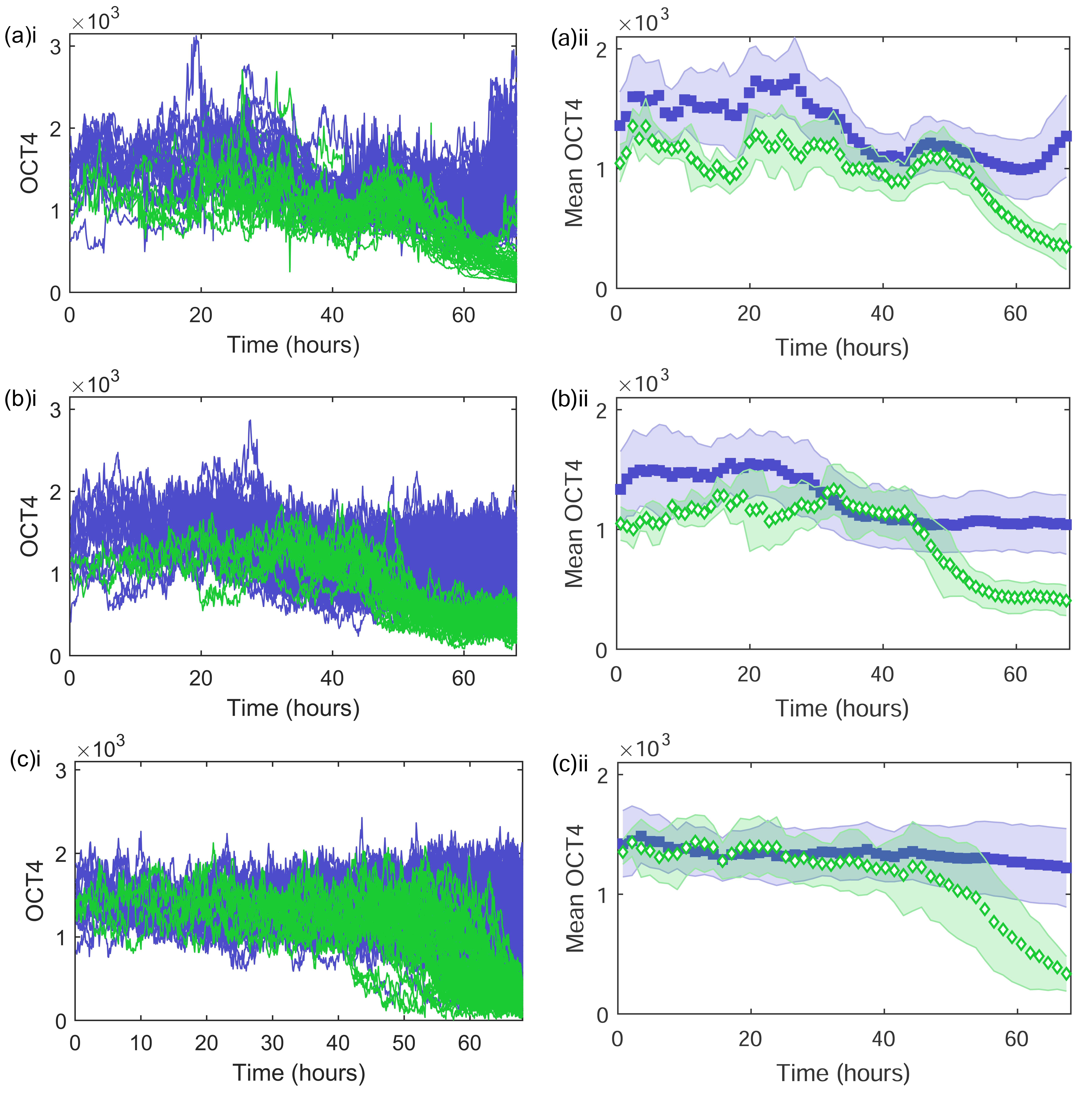}
	\caption{\label{fig:diff_sims} {\bf The experimental and simulated dynamics of OCT4 upon differentiation at 40 hours.} The (a) experimental (i) OCT4 and (ii) mean OCT4 with time. The (b) simulated (i) OCT4 and (ii) mean OCT4 with time with differentiation induced at 40 hours using a time-dependent carrying capacity, Eq~(\ref{eq:timedepk}), with the parameters specified in Table~\ref{tab:sim_parameters_diff}. The (c) simulated (i) OCT4 and (ii) mean OCT4 wth time with differentiation induced at 40 hours by introducing an Allee effect term to the SLE, Eq~(\ref{eq:SLEallee}), with $r=0.025$, $K=1290$, $\sigma_{\rm{A}}=35$, $\sigma_{\rm{M}}=0.035$ and $A=1000$.}
\end{figure}

In the previous section we considered modelling temporal OCT4 regulation before any differentiation stimulus (BMP4) is added, corresponding to the time interval $0<t<40\,$h in the experimental colony \cite{Wolff18,Wadkin20oct}. The addition of BMP4 causes a significant reduction in OCT4 expression in the differentiated cells, shown in Fig~\ref{fig:octovertime}(a). The mean OCT4, shown in Fig~\ref{fig:diff_sims}(a) also shows the clear reduction in differentiated cells. The median and mode experimental OCT4 are shown in \nameref{medmode}. We explore two methods of modelling this reduction in OCT4 as differentiation is induced. Firstly, we apply the SLE with a time-dependent carrying capacity as discussed previously, and secondly, we consider the use of the SLE with an Allee effect. Although not seen in this experiment, it should be noted that high OCT4 values can also correspond to cell differentiation \cite{Strebinger19}.

\subsubsection*{Differentiation with a time-dependent carrying capacity}

We previously employed the SLE with a time-dependent carrying capacity, Eq~(\ref{eq:timedepk}), to simulate a moderate reduction in the average OCT4 expression post-25 hours, as shown in Fig~\ref{fig:shiftk}. We could extend this technique to simulate the more drastic reduction in OCT4 seen when the differentiation stimulus is added. 

As before, we can estimate the carrying capacities for $t<25\,$h as $K_{\rm{p}}\approx1500$ and $K_{\rm{d}}\approx1100$ for pluripotent and differentiated cells, respectively. For $t>25\,$h we can simulate the modest reduction in OCT4 expression for the pluripotent cells with a reduction of the carrying capacity to $K_{\rm{p}}\approx1000$. For the differentiated cells, a reduction to $K_{\rm{d}}\approx300$ in the time interval $t>40\,$h corresponds to cell differentiation. These shifting carrying capacities, along with the other model parameters are given in Table~\ref{tab:sim_parameters_diff}. The dynamics under this regime are shown in Fig~\ref{fig:diff_sims}(b) and \nameref{medmode}. The time-dependent carrying capacity leads to the reduction of OCT4 in the differentiated cell group, well capturing the dynamics of the experiment. 

\begin{table}[!h]
	\centering
		\caption{Simulation parameters for the OCT4 expression of pluripotent and differentiated cells using the SLE with additive and multiplicative noise, and a time-dependent carrying capacity, Eq~(\ref{eq:timedepk}), to capture induced differentiation.}
	\begin{tabular}{@{}lcccc@{}}
		\toprule
		& Parameter & $0\leq t<25\,$h & $25\leq t<40\,$h & $40\leq t<68\,$h\\
		\midrule
		\multirow{5}{*}{Pluripotent} & $N_0$ & \multicolumn{3}{c}{14} \\
		& $r$ &  \multicolumn{3}{c}{0.015}  \\
		& $K$ & 1500 & 1000 & 1000 \\
		& $\sigma_{\rm{A}}$ &  \multicolumn{3}{c}{35}  \\
		& $\sigma_{\rm{M}}$ &  \multicolumn{3}{c}{0.035} \\
		& $H$ & \multicolumn{3}{c}{0.38}\\
		\midrule
		\multirow{5}{*}{Differentiated} & $N_0$ &\multicolumn{3}{c}{2} \\
		& $r$ &  0.015 & 0.015 & 0.008 \\
		& $K$ & 1100 & 1100 & 300 \\
		& $\sigma_{\rm{A}}$ &  \multicolumn{3}{c}{25} \\
		& $\sigma_{\rm{M}}$ & \multicolumn{3}{c}{0.03}  \\
		& $H$ & \multicolumn{3}{c}{0.38} \\
		\bottomrule
	\end{tabular}
	\label{tab:sim_parameters_diff}
\end{table}

This model could be further refined by the use of a more sophisticated function for the time-dependent carrying capacity, which could be estimated from experimental data such as that in Ref.~\cite{Strebinger19} and \cite{Wolff18}. The model could also easily be adapted to include a population of cells exhibiting high OCT4 values pre-differentiation, with a corresponding increase in their carrying capacity. However, the model would remain purely descriptive, with pro-pluripotent and pro-differentiated cells defined from the outset with different behavioural rules. Next we consider using the SLE with an Allee effect to simulate differentiation and identify the different cell fate types.

\subsubsection*{Differentiation with an Allee effect}

Another possible method of modelling induced differentiation is the SLE with a demographic Allee effect. Allee effects are traditionally used for modelling population numbers, with the effect inhibiting population growth at low densities as observed in both animal and cell populations \cite{Drake11,Gascoigne04,Johnson19}. The deterministic logistic equation for OCT4 expression $O$ with this effect incorporated has the form
\begin{equation}
\frac{dO}{dt}=rN\left(1-\frac{O}{K}\right)\left(\frac{O-A}{K}\right),
\label{eq:alleedet}
\end{equation}
where $A$ is critical point at which the Allee effect occurs. Note that there are other methods of simulating Allee effects through e.g., difference equations \cite{Saber10,Wang02} and Lotka-Voltera models \cite{Zhou05,Lin18}. Here we use the logistic equation for consistency with our previous modelling results. 

The effect of the Allee term in Eq~(\ref{eq:alleedet}) on both $dO/dt$ and the OCT4 expression $O$ for an example system is illustrated in \nameref{allee_eg}. For a weak Allee effect, $A<O(t=0)$, the rate of change $dO/dt$ remains positive for $O<K$ but is significantly suppressed. For a stonger Allee effect, $A>O(t=0)$, $dO/dt$ is negative for $O<K$ and results in the OCT4 expression declining to zero. It is this declining effect we can employ to simulate the reduction in OCT4 expression for the differentiated cells. The Allee effect can be introduced at a certain time point resulting in either continued suppressed growth or a decline to zero. Examples of `switching on' both weak and strong Allee effects during logistic growth are shown in \nameref{allee_eg2}.

For simulating OCT4 expression through the differentiation process with the SLE, we can switch on the Allee effect term at the time the differentiation agent is added (40\,h). If the OCT4 expression is below $A$, then the Allee effect will be strong and the OCT4 will decline to zero. The stochasticity in the system will mean that only some of the cells will meet this condition, with others having an OCT4 expression greater than $A$, and therefore continuing with (suppressed) logistic growth. The stochasticity will also result in this effect taking place at all times past $40\,$h, so the differentiation process will happen at different times for different cells. The SLE for $X=\ln(O)$ with additive fBm noise $\xi_1$ and multiplicative fBm noise $\xi_2$ is 
\begin{equation}
\frac{dX}{dt}=r\left(1-\frac{e^X}{K}\right)\left(\frac{e^X-A}{K}\right)+\frac{\sigma_{\rm{A}}}{e^X}\xi_1+\sigma_{\rm{M}}\xi_2,
\label{eq:SLEallee}
\end{equation}
where $A$ is the Allee effect critical point. 

The OCT4 dynamics for 16 cells simulated with the SLE, Eq~(\ref{eq:stochlog2bothnoise}), for $t<40\,$h and the SLE with an Allee effect, Eq~(\ref{eq:SLEallee}), for $t\geq 40\,$h with $r=0.025$, $K=1290$, $\sigma_{\rm{A}}=35$, $\sigma_{\rm{M}}=0.035$ and $A=1000$ are shown in Fig~\ref{fig:diff_sims}(c) and \nameref{allee_diff2}. Here the fates of each cell are identified at the end of the simulation, with the cells whose OCT4 has reduced as a result of the Allee effect classed as differentiated, and the cells whose OCT4 has remained constant as pluripotent. The model captures the reduction of OCT4 in the differentiated subset of cells whilst keeping a remaining pluripotent cell population. However, the OCT4 in the pro-differentiated group pre-Allee effect is no lower than for the pro-pluripotent cell group, unlike in the experimental results. Furthermore, an additional model would be required to introduce differentiated cells with high OCT4 values.

\section*{Discussion}
We have explored different modelling techniques for describing temporal OCT4 regulation, guided by previous analysis of experimental OCT4 expression in a growing hESC colony \cite{Wolff18,Wadkin20oct}, particularly fractional Brownian motion and the stochastic logistic equation. A differentiation agent, BMP4, was added to the cells at 40 hours and results in the reduction of OCT4 expression in the differentiated cells. Although not seen here, it is also possible for high OCT4 expression to accompany cell differentiation \cite{Strebinger19}. Pre-BMP4 we identified some key features including an anti-persistent stochasticity, and for pluripotent cells a temporal skew and shifting mode in the distribution of all OCT4 expressions. All the models discussed follow a common base model which sets up the initial conditions and describes cell proliferation. When adjusted to produce unsynchronised cell divisions, the base model describes well the population growth with time, shown in Fig~\ref{fig:basemodel}(d). We then focus on different mathematical methods of generating the temporal OCT4 expressions for the cell population within this base model. The simulated populations consist of 16 cells (with 14 pro-pluripotent and two pro-differentiated) resulting in approximately 30000 simulated OCT4 expressions.  We have taken a systematic approach, gradually building complexity to illustrate the methodology of developing statistical models for biological systems.  

Firstly, we consider modelling the OCT4 dynamics pre-BMP4, i.e., for $t<40\,$hours. The analysis in Ref. \cite{Wadkin20oct} revealed that OCT4 values fluctuate stochastically with anti-persistence and a Hurst exponent of 0.38, suggesting the use of fractional Brownian motion (fBm) \cite{Mandelbrot68}. There is also further experimental evidence that gene expressions and transcription factor dynamics display fractal characteristics \cite{Ghorbani18}. The use of fBm is particularly common in financial modelling \cite{Cheridito03,Xiao10,Bender11}, but it has also been used to describe diffusion within crowded fluids (such as the cytoplasm of cells) \cite{Ernst12} and the kinetics of transcription factors \cite{Woringer20}. The stochasticity from fBm results in a wider range of OCT4 values at later times than seen experimentally (an effect which is exacerbated with time). 

The range of OCT4 can be controlled artificially with boundary conditions (either absorbing or reflecting), but the overall distribution of all OCT4 values is not well captured, shown in \nameref{fbmsim}. It is also unclear whether these boundary conditions are biologically appropriate as OCT4 expression is regulated by a complex range of factors across the transcriptional, post-transcriptional and epigenetic regulation levels \cite{Li04,Wang12,Shi10,Babaie07}. Interestingly, mechanical limits to transcription have been shown to naturally generate bounds to transcriptional noise \cite{Sevier16}. A boundary condition at zero corresponds to the fact that OCT4 expression never becomes negative with the upper boundary representing a maximum possible value. Furthermore, what is the biological implication of the removal of cells through through absorbing boundaries or the recovery of expression through reflecting boundaries? One possibility for absorbing boundaries for pro-pluripotent cells is to represent differentiation happening at both the upper and lower boundary \cite{Strebinger19}. Although fBm alone is not sufficient to capture the experimental behaviour, it does (by design) capture the anti-persistence ($H=0.38$) and so in all later model iterations we use fBm noise to generate the stochasticity. 

A somewhat less artificial method of keeping the OCT4 values within range is to use the stochastic logistic equation (SLE), which has a regulating parameter of the carrying capacity, $K$, which represents the maximum amount of OCT4 that can be expressed within each individual cell. Note that this could be due to limits on the expression of OCT4 due to other members of the regulatory network which cause its down-regulation. In our model, the stochasticity allows for some fluctuations above $K$. Similarly to the boundary conditions this maximum value depends on the complex inter-regulatory network of OCT4, however, we estimate the value of the carrying capacity from the experimental results as the median of all OCT4 values (taking into account the stochasticity allowing for $O>K$). 

There are many sources of noise within the system, with internal noise resulting from stochastic chemical reactions (represented by additive noise) and external noise originating from fluctuations in other cellular components that indirectly cause variation in transcription factor dynamics \cite{liu2009effect}. We consider both additive and multiplicative noise, shown in Fig~\ref{fig:sle_addnoise}. The introduction of multiplicative noise creates larger fluctuations above the carrying capacity, qualitatively similar to those seen in the experiment. This results in a distribution of all OCT4 values well matched to the experiment, with the slight positive skew being captured. Both additive and multiplicative noise can be used to regulate gene expression, with multiplicative noise allowing small deviations in transcription rates to lead to large fluctuations in protein productions  \cite{Hasty00}. 

A property not captured by the SLE with either additive or multiplicative noise is the time-dependency of this positive skew. It occurs only at later times, and only in pluripotent cells, shown in the time-discretised distributions of OCT4 in Fig~\ref{fig:octovertime}(c). This temporal skew can be captured by the SLE with both additive and multiplicative noise, with the type of noise time-dependent; additive noise at early times produces symmetrical distributions of OCT4, with multiplicative noise at later times producing skewed distributions, shown in Fig~\ref{fig:noiseswitch}. Here we changed the noise function stepwise, but this could be further smoothed using a more sophisticated time-dependent noise function.

Another interesting property of the experimental OCT4 is the decline in expression for pluripotent cells post-25\,hours, shown in Fig~\ref{fig:octovertime}(c). We consider capturing this behaviour using the SLE with a time-dependent carrying capacity. Since this parameter is likely to depend on a large number of biological factors, it is not unreasonable to expect that it may change with environmental conditions and experimental time. We consider the pluripotent and differentiated cells separately, each with a different carrying capacity, corresponding to the suggestion that the decision to differentiate is determined pre-differentiation stimulus \cite{Wolff18}. The carrying capacity for both cell groups is reduced at 25 hours, resulting in a decline in OCT4 expression, particularly for the pluripotent cell group with originally higher expression. Although this technique well describes the experimental results (shown in Fig~\ref{fig:shiftk}), it requires multiple parameters which need to be elucidated from further experimental data.

We then consider modelling the OCT4 regulation for all times, including the decline in expression due to the addition of the differentiation stimulus. We extend the time-dependent carrying capacity approach, reducing the carrying capacity further for the differentiated cell group at 40 hours. This well captures the decline in OCT4 upon differentiation, along with the more subtle decline in pluripotent cells, shown in Fig~\ref{fig:diff_sims}(b). Here we have used a stepwise change in the parameter $K$, but this is easily adjustable to other experimental results and more sophisticated functions could be used to capture other trends. Similarly, a population of high OCT4 differentiated cells could be introduced with a corresponding increase in their carrying capacity. The pro-differentiated cells are identified from the outset and although this is not biologically unreasonable, with evidence that cell fate is determined pre-differentiation agent \cite{Wolff18}, the model itself does not produce the two fate groups which limits its future capacity to develop into a predictive model. 

A method of inducing differentiation which naturally produces the two fate groups is the SLE with an Allee effect. Allee effects are well used across mathematical biology \cite{Drake11,Gascoigne04,Johnson19}, but we are not aware of their application to pluripotency transcription factor expression. The Allee effect results in a decline to zero for cells whose OCT4 expression fluctuates below the critical point $A$. The stochasticity in the system means that this condition is met for only some of the cells, causing the formation of a differentiated cell group with reducing or zero OCT4 and a pluripotent cell group with stable OCT4 expression at the carrying capacity, shown in Fig~\ref{fig:diff_sims}(c). This model is limited to describing low OCT4 differentiated cells as seen in this experiment and high OCT4 differentiation would need to be incorporated through another technique. This model could be combined with a time-dependent carrying capacity to capture the decline in expression in pluripotent cells. 

The models discussed here are of a purely descriptive nature, but outline a possible framework for modelling the regulation of OCT4. We have explored systematically a wide range of effects that might be able to reproduce rather fine details in the experimentally observed dynamics of the OCT4 expression and identified an adequate and optimal combination of such effects. However, the resulting model may not be unique and other approaches may be viable. To justify any model of this kind and to develop it into a prognostic tool for \textsl{in-silico} experimentation, it should be assessed and compared with targeted experiments. With this caveat, we believe that the model developed can be used as a provisional prognostic tool and basis for further mathematical model development. A summary of the models discussed and the experimental properties they capture corresponding to the key features identified is given in Table~\ref{tab:models}. Further time-lapse experiments monitoring PTFs are needed to confirm which of these properties are inherent to OCT4 expression, and how they vary depending on experimental conditions, and to provide more extensive benchmarking for the modelling approaches and assumptions. It will be informative to apply the same quantitative framework to the other predominant transcription factors, SOX2 and NANOG. Their individual regulatory dynamics could then be compared using the key descriptive parameters, and any systematic differences identified. This information will help build the picture of the wider PTF system with the dynamics of the PTFs considered as part of an inter-linked network. In general, this highlights the need for further temporal experimental data on PTF regulation to build upon this mathematical framework and develop more sophisticated predictive models. These models of the microstate of PTF regulation will help inform longer time-scale models of the pluripotent macrostate.

\begin{table}[!h]
	\begin{adjustwidth}{-2.25in}{0in}
	\caption{A summary of the key features identified experimentally and the models used to describe each behaviour.}
	\begin{tabular}{@{}p{2.75cm}p{9cm}p{6cm}@{}}
		\toprule
		& Key features & Model\\
		\midrule
		\multirow{4}{*}{Pre-differentiation} & 1. Stochastic noise with Hurst exponent of 0.38. & fBm, Eq~\ref{eq:fbm}\\
		& 2. Pro-differentiated cells show reduced OCT4 throughout. & Incorporated through initial conditions.\\ 
		& 3. Positive skew of all pro-pluripotent OCT4 expressions. & SLE (multiplicative noise), Eq~\ref{eq:stochlog2mult} and \ref{eq:stochlog2bothnoise}\\ 
		& 4. Reduction in pro-pluripotent OCT4 post 25 hours. & SLE ($K(t)$), Eq~\ref{eq:timedepk}\\
		\midrule
		\multirow{2}{*}{Post-differentiation} & 1. Reduction in OCT4 expression for some cells. & SLE ($K(t)$ or Allee effect), Eq~\ref{eq:timedepk} or Eq~\ref{eq:SLEallee}\\
		& 2. Separation into pluripotent and differentiated groups. & SLE (Allee effect), Eq~\ref{eq:SLEallee}\\ 
		\bottomrule
	\end{tabular}
	\label{tab:models}
	\end{adjustwidth}
\end{table}

\section*{Acknowledgments}
LEW would like to acknowledge support from the London Mathematical Society (Early Career Fellowship). ML acknowledges BBSRC UK (BB/I020209/1). IN acknowledges the
grant from the Russian Government 641 Program for the recruitment of the leading scientists into 641 Russian Institution of Higher Education 14.w03.31.0029
and RFFI project grant number 20-015-00060.

\nolinenumbers

%
%
%

\clearpage
\newpage

\bibliography{refs}

\clearpage
\newpage
\section*{Supporting information}

\renewcommand\thefigure{S\arabic{figure}}    
\setcounter{figure}{0}

\paragraph*{S1 Fig.}
\label{fbm}
{\bf The effect of the Hurst exponent.} Realisations of simulated noise in fractional Brownian motion with (a) $H=0.1$ (anti-persistence), (b) $H=0.5$ (Brownian) and (c) $H=0.9$ (persistence), (d-f) the corresponding simulated trajectories with initial condition $B_H(0)=0$. (g-i) Simulation of OCT4 for 40 hours, with ten initial cells, and temporal OCT4 determined by simulated realisations of $\sigma B_H$ with $\sigma=90$ and (g) $H=0.1$, (h) $H=0.5$ and (i) $H=0.9$.

\paragraph*{S2 Fig.}
\label{fbmsim}
{\bf The effect of boundary conditions on simulated OCT4.} Simulated OCT4 expression (blue) using fBm with 16 initial cells, $\sigma=90$ and $H=0.38$ with (a) no, (c) absorbing, and (e) reflecting boundary conditions at zero and 2500. The experimental data is overlaid in orange. The corresponding histograms for simulated OCT4 (blue) with (b) no, (d) absorbing, and (f) reflecting boundaries. The kernel density fitting to the experimental distribution is shown in orange.

\paragraph*{S3 Fig.}
\label{loglogisticeg}
{\bf The SLE with multiplicative noise.} Realisations of the dynamics of (a) $X=\log(O)$ and (b) $O=e^X$ from Eq~(\ref{eq:stochlog2mult}) with $r=0.1$/h, $K=100$, $\xi=W_t$ and $\sigma_{\rm{M}}=0$ (blue), 0.025 (orange) and 0.075 (yellow).

\paragraph*{S4 Fig.}
\label{medmode}
{\bf The average experimental and simulated OCT4 expressions using a time-dependent carrying capacity.} The (a,c) experimental and (b,d) simulated median and mode OCT4. The dynamics are simulated using the SLE with additive and multiplicative noise, and a time-dependent carrying capacity, Eq~(\ref{eq:timedepk}), with parameters specified in Table~\ref{tab:sim_parameters_diff}.

\paragraph*{S5 Fig.}
\label{allee_eg}
{\bf The deterministic logistic equation with a demographic Allee effect.} The deterministic logistic equation with an initial condition of $O_0=10$, $r=0.1\,$/h, $K=50$ and an Allee effect, Eq~(\ref{eq:alleedet}), for (a) $dO/dt$ and (b) $O$ with $A=1$ (orange) and $A=50$ (green). The deterministic logistic equation with no Allee effect is shown in blue.

\paragraph*{S6 Fig.}
\label{allee_eg2}
{\bf Switching on a demographic Allee effect.}  The deterministic logistic equation with an initial population size of $N_0=10$, $r=0.1\,$/h and $K=50$ (blue). The Allee effect term in Eq~(\ref{eq:alleedet}) is introduced at $t=25\,$h with (a) $A=20$ (orange) and $A=25$ (green) and (b) $A=40$ (orange) and $A=50$ (green). The deterministic logistic growth with no Allee effect is shown as blue dashed.

\paragraph*{S7 Fig.}
\label{allee_diff2}
{\bf  The average experimental and simulated OCT4 expressions using an Allee effect.} The (a,c) experimental and (b,d) simulated median and mode OCT4. The dynamics are simulated using the SLE with an Allee effect at 40 hours, Eq~(\ref{eq:SLEallee}), with $r=0.025$, $K=1290$, $\sigma_{\rm{A}}=35$, $\sigma_{\rm{M}}=0.035$ and $A=1000$.

\newpage
\begin{figure}[!h]
	\includegraphics[width=\textwidth]{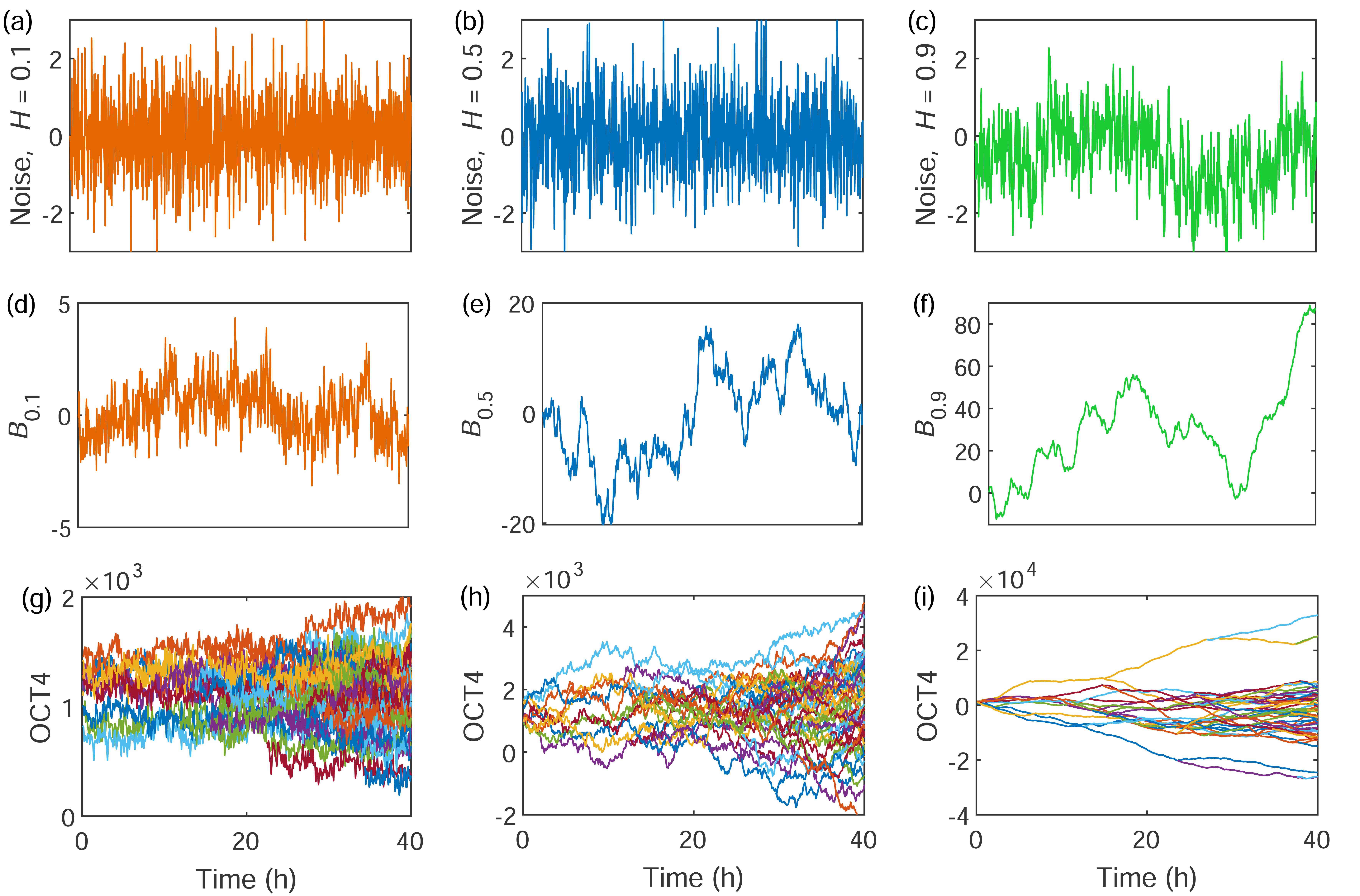}
	\caption{\label{fig:fbm}}
\end{figure}

\begin{figure}[!h]
	\includegraphics[width=\textwidth]{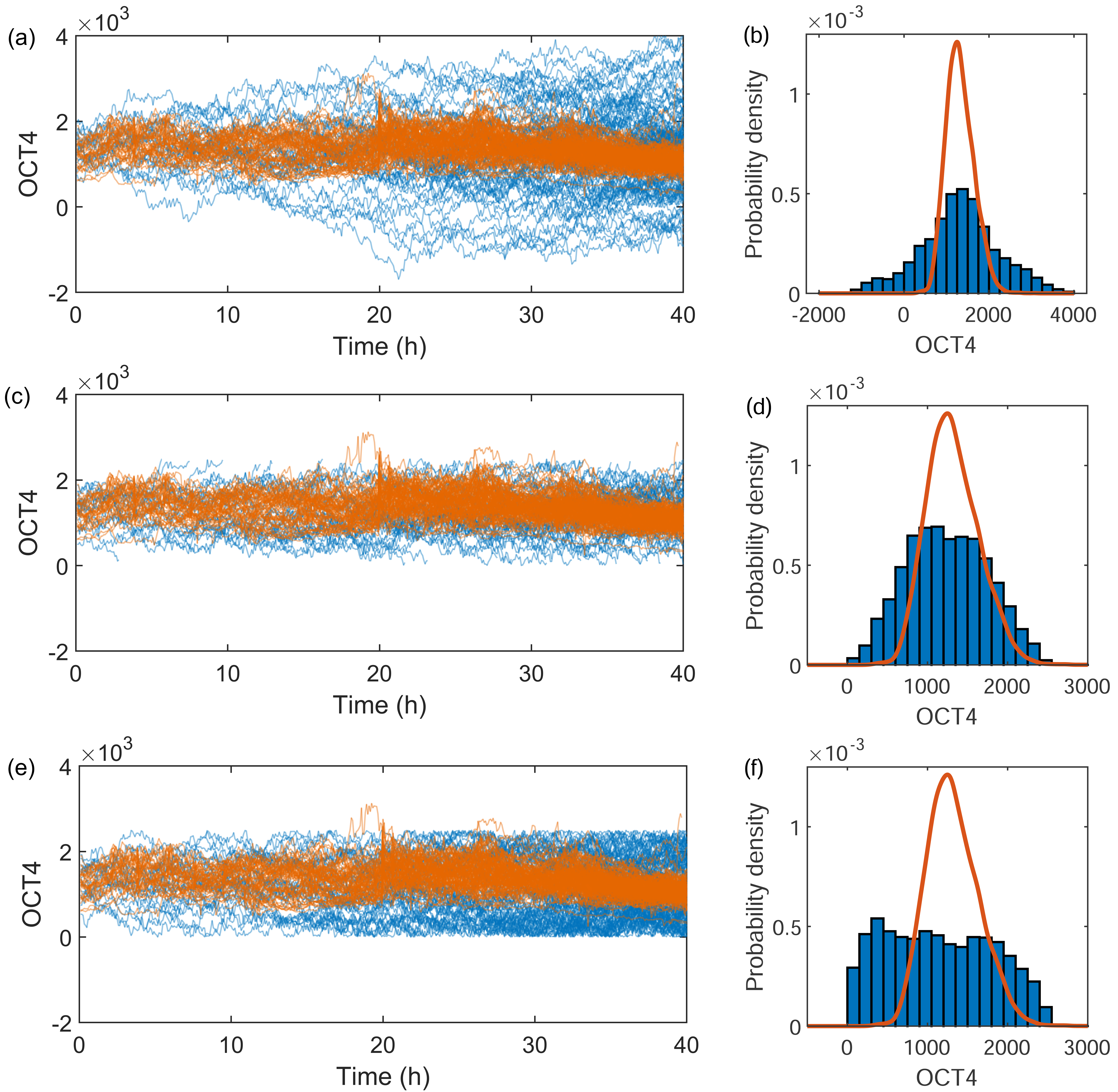}
	\caption{\label{fig:fbmsim}}
\end{figure}

\begin{figure}[!h]
	\includegraphics[width=\textwidth]{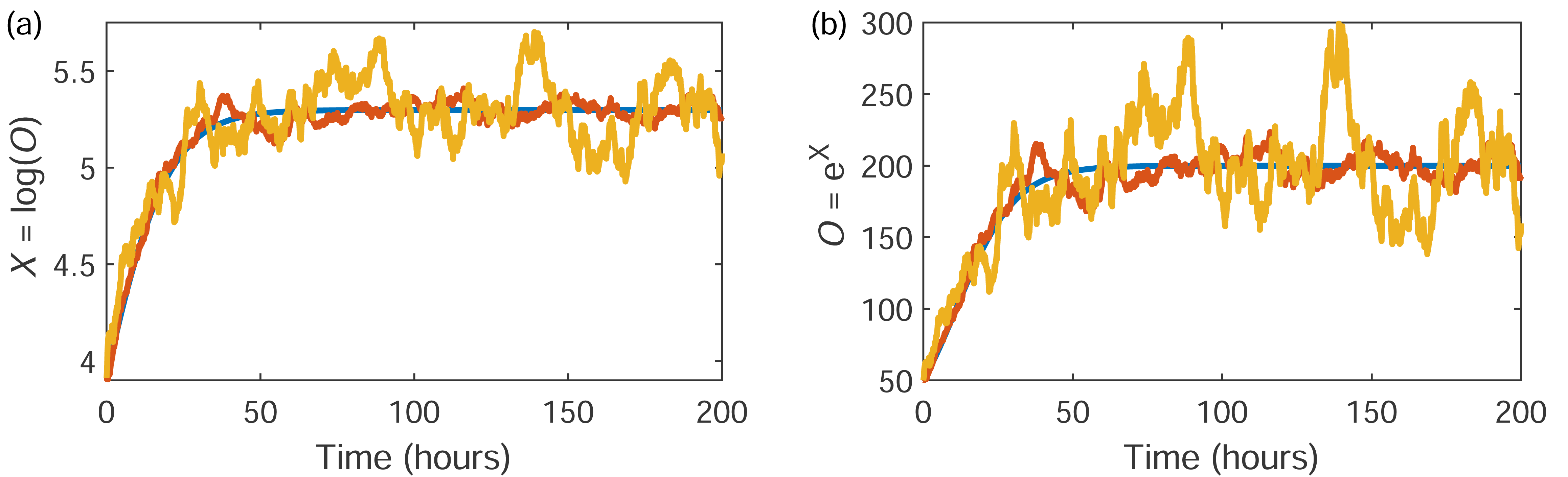}
	\caption{\label{fig:loglogisticeg}} 
\end{figure}

\begin{figure}[!h]
	\includegraphics[width=\textwidth]{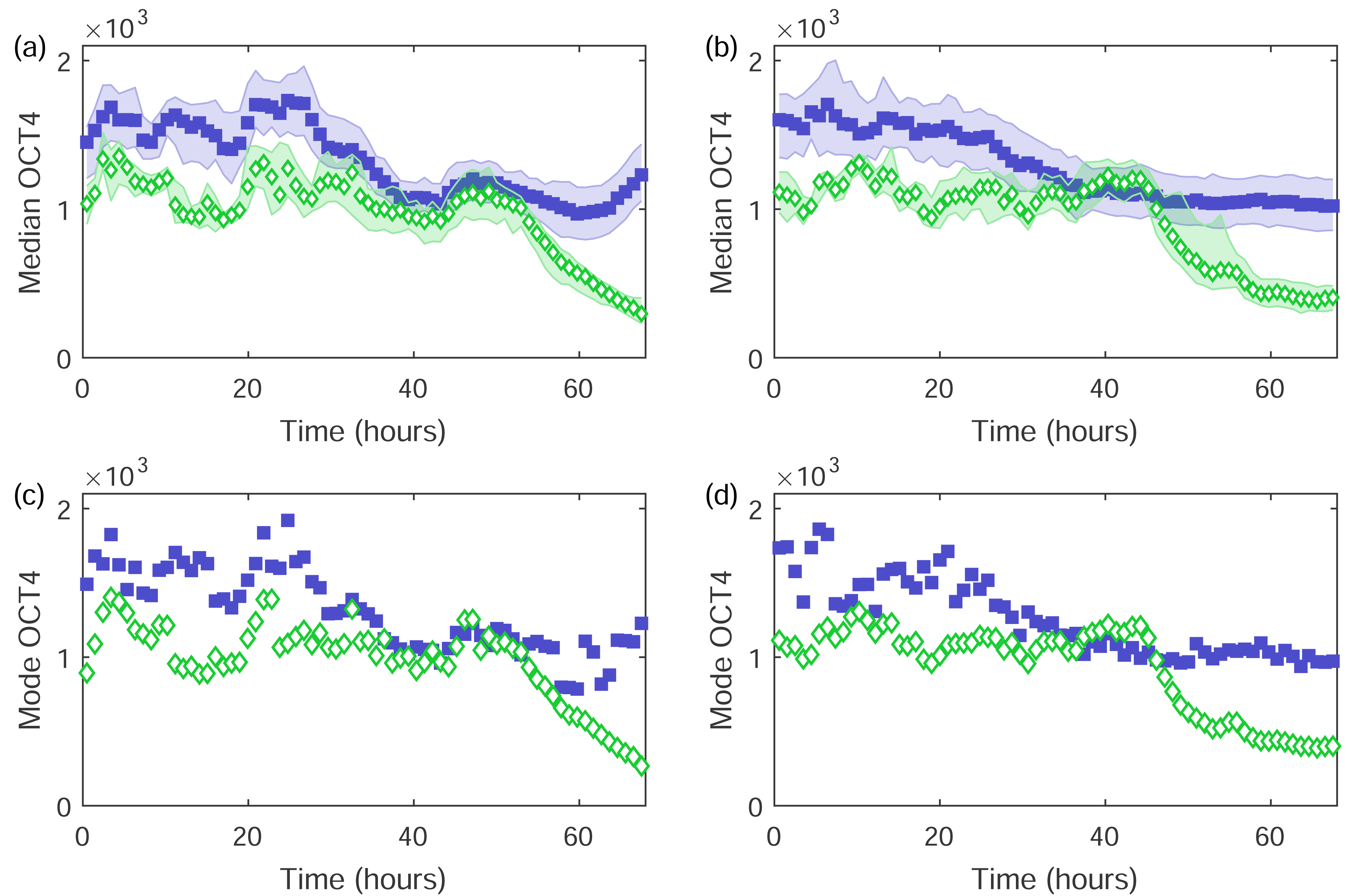}
	\caption{\label{fig:medmode}}
\end{figure}

\begin{figure}[!h]
	\includegraphics[width=\textwidth]{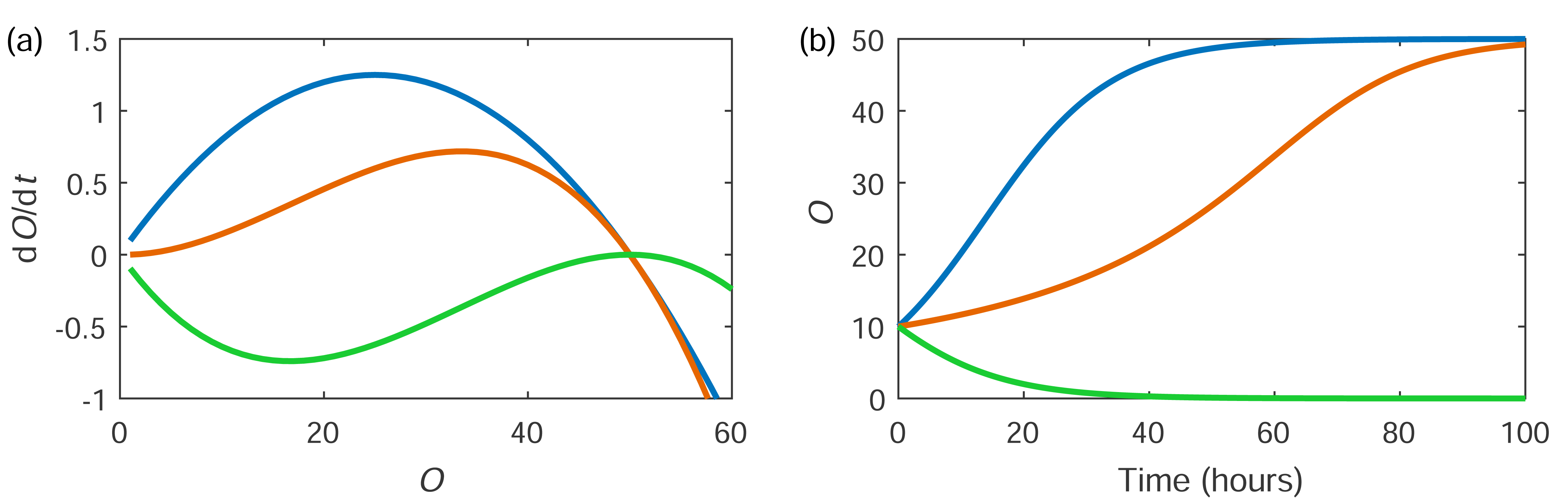}
	\caption{\label{fig:allee_eg} }
\end{figure}

\begin{figure}[!h]
	\includegraphics[width=\textwidth]{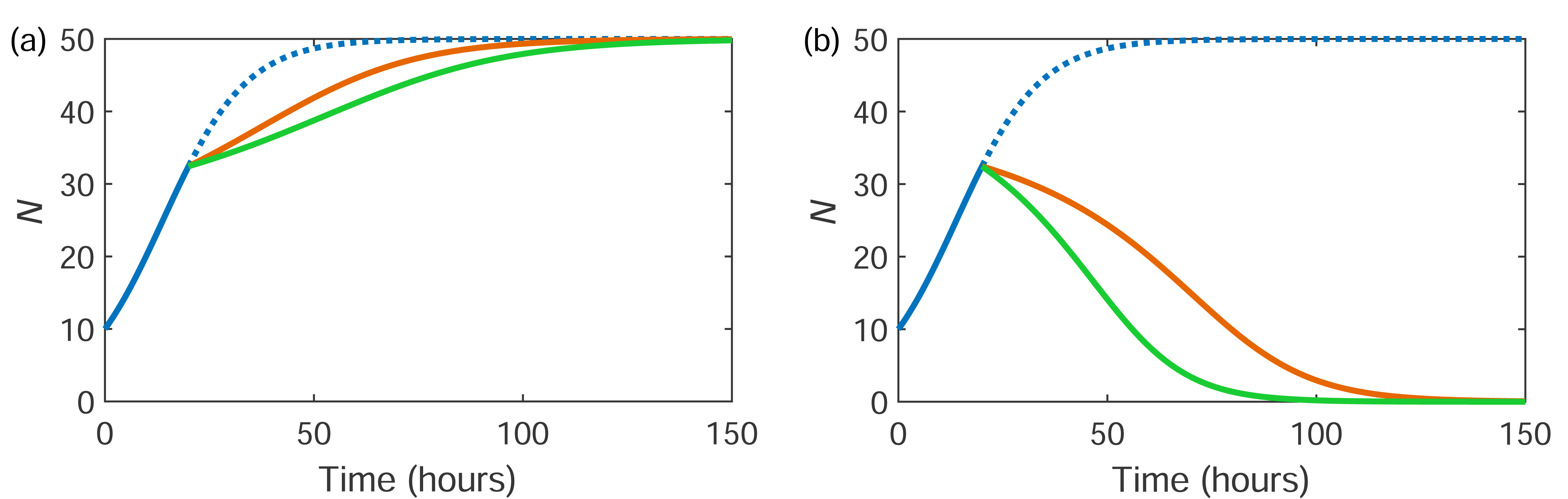}
	\caption{\label{fig:allee_eg2}}
\end{figure}

\begin{figure}[!h]
	\includegraphics[width=\textwidth]{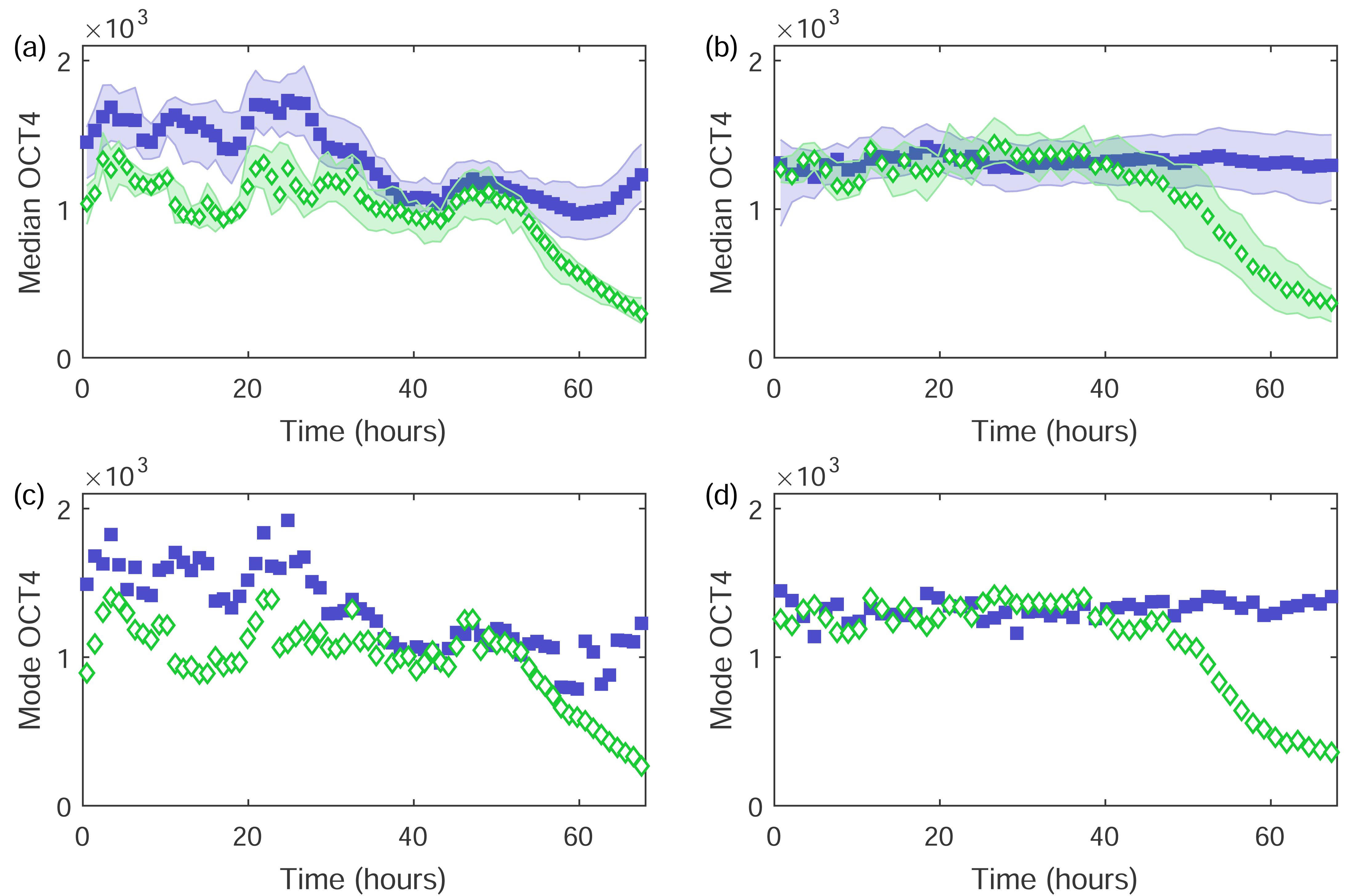}
	\caption{\label{fig:allee_diff2}}
\end{figure}

\end{document}